\documentclass[a4paper,11pt]{article}
\usepackage{pos}
\usepackage{overpic}
\usepackage{subfigure}
\usepackage{multirow}
\usepackage{booktabs}
\usepackage{tabularx}

\title{Measurements of charmonia decays from BESIII}

\author*{Han Miao}
\author{on behalf of the BESIII collaboration}

\affiliation{Institute of High Energy Physics,\\
  19B Yuquan Road, Shijingshan District, Beijing 100049, People's Republic of China}

\emailAdd{miaohan@ihep.ac.cn}

\abstract{In this talk, recent measurements of charmonium decays of BESIII are presented. Using 448 million $\psi(3686)$ events collected with the BESIII detector, the branching fractions of the decays $\chi_{cJ} \to \phi \phi (J=0,1,2)$ have been measured most precisely, and the polarization parameters of $\chi_{cJ} \to \phi \phi$ have been determined for the first time via a helicity amplitude analysis. Using the same data sample as in the previous study, first evidence of $\eta_c(2S) \to \pi^+ \pi^- \eta$ has been found in the decay sequence $\psi(3686) \to \gamma \eta_c(2S)(\to \pi^+ \pi^- \eta)$. The product of the branching fractions of $\psi(3686) \to \gamma \eta_c(2S)$ and $\eta_c(2S) \to \pi^+ \pi^- \eta$ is reported as well as the individual branching fraction of $\eta_c(2S) \to \pi^+ \pi^- \eta$. The process $e^+ e^- \to \eta J/\psi$ at a center-of-mass energy $3.773~{\rm GeV}$ is observed for the first time. Its Born cross-section is measured, and the branching fraction of $\psi(3770) \to \eta J/\psi$ is determined by a combined fit with the cross sections at other energy points, after considering the interference effect for the first time. Utilizing 2708 million $\psi(3686)$ events collected by the BESIII detector, the decays $\chi_{cJ} \to \Omega^- \bar{\Omega}^+(J=0,1,2)$ have been observed for the first time with high significance via the radiative decays of $\psi(3686) \to \gamma \chi_{cJ}$. The relevant branching fractions have been provided.}

\FullConference{21st Conference on Flavor Physics and CP Violation (FPCP 2023)\\
 29 May - 2 June 2023\\
 Lyon, France\\}

\begin{document}
\maketitle

\section{Introduction}

Since the discovery of $J/\psi$ in the winter of 1974, heavy quarkonia have always been an ideal field for physicists to study the main properties of quantum chromo-dynamics (QCD)~\cite{Kwong:1987mj, QuarkoniumWorkingGroup:2004kpm, Eichten:2007qx, Brambilla:2010cs, Rosner:2011eg}, among which charmonia, especially, play an important role for understanding the physics in the energy region between perturbative and non-perturbative QCD. Clear spectrum of low-lying charmonium states below the open-charm $D\bar{D}$ threshold has been observed experimentally and predicted theoretically within the potential models, which incorporate a color Coulomb term at short distances and a linear scalar confining term at large distances. Therefore, for a long time, the charmonium system has become the prototypical ``hydrogen atom'' of meson spectroscopy, while the study of charmonium decays to light hadrons has always sufferred from the difficulties of non-perturbative calculation. In history, multiple models or techniques are raised for this issue, of which none describes all the experimental measurements perfectly. The case will be much more complex if considering the decays to light hadrons, thus the experimental results are really essential for the research on charmonium decays.

In the experimental aspect, charmonium states have been studied thoroughly by CLEO-c, BES and other charm or bottom factories in history, and are still charming after nearly 50 years with the new decay channels and new states continuously discovered. As the only facility running in the tau-charm energy region around world now, BEPCII and BESIII dedicate massive investigation, arousing widespread interest for theoretical studies. The following part will be a brief introduction and summary of the recent results on the measurement of charmonium decays at BESIII.

\section{BEPCII and BESIII}

\begin{figure}[htbp]
	\centering
	\includegraphics[width=0.7\textwidth]{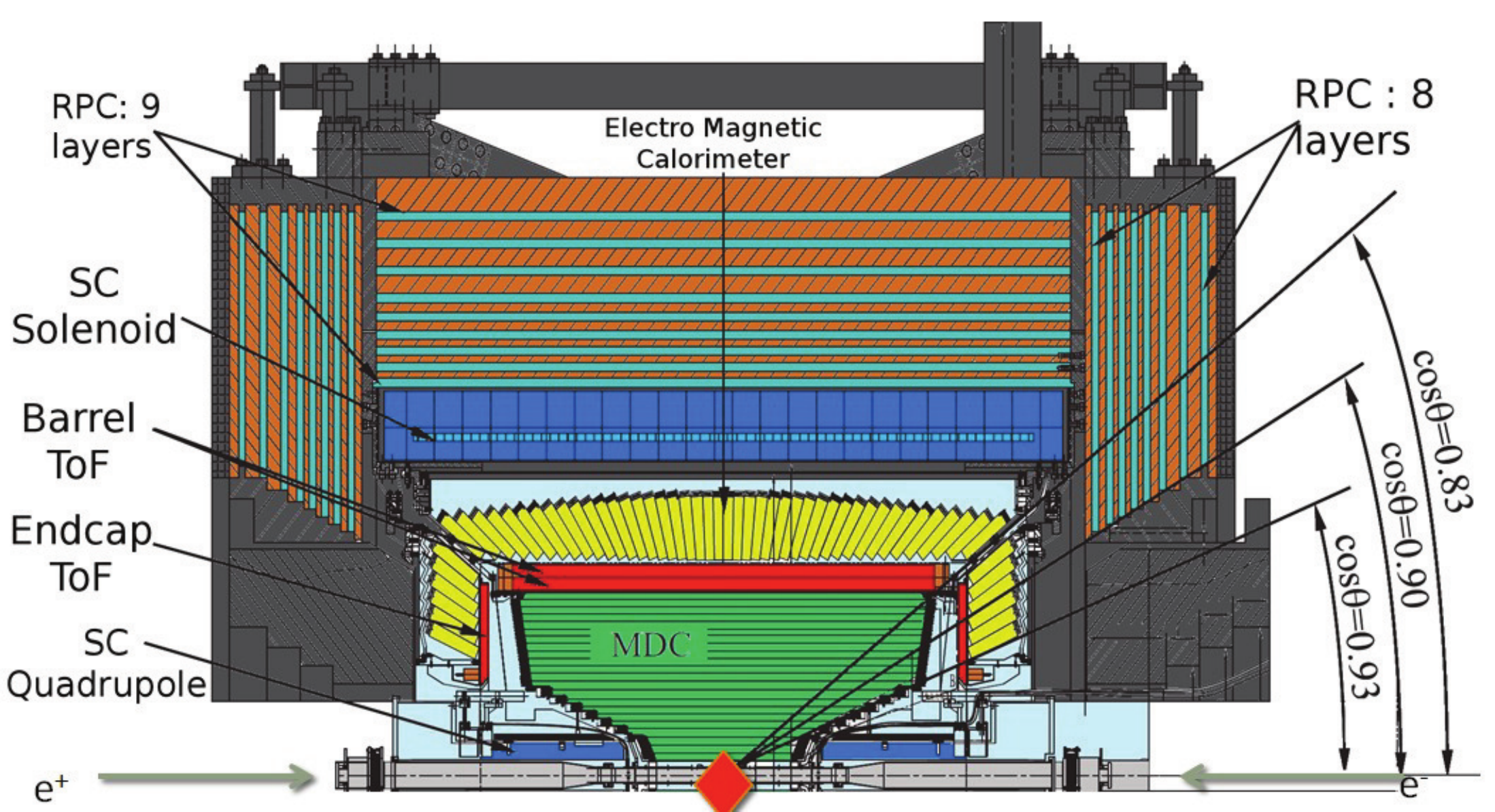}
	\caption{The BESIII detector.}
	\label{fig:BESIII}
\end{figure}

The BESIII detector~\cite{Ablikim:2009aa}, as shown in Figure~\ref{fig:BESIII}, records symmetric $e^+e^-$ collisions provided by the BEPCII storage ring~\cite{Yu:IPAC2016-TUYA01} in the center-of-mass energy range from 2.0 to 4.95~GeV, with a peak luminosity of $1.1 \times 10^{33}\;\text{cm}^{-2}\text{s}^{-1}$ achieved at $\sqrt{s} = 3.77~\text{GeV}$. BESIII has collected large data samples in this energy region~\cite{Ablikim:2019hff, EcmsMea, EventFilter}. The cylindrical core of the BESIII detector covers 93\% of the full solid angle and consists of a helium-based multilayer drift chamber~(MDC), a plastic scintillator time-of-flight system~(TOF), and a CsI(Tl) electromagnetic calorimeter~(EMC), which are all enclosed in a superconducting solenoidal magnet providing a 1.0~T magnetic field. The magnetic field was 0.9~T in 2012, which affects 10.8\% of the total $J/\psi$ data. The solenoid is supported by an octagonal flux-return yoke with resistive plate counter muon identification modules interleaved with steel. The acceptance of charged particles and photons is 93\% over $4\pi$ solid angle. The charged-particle momentum resolution at $1~{\rm GeV}/c$ is $0.5\%$, and the ${\rm d}E/{\rm d}x$ resolution is $6\%$ for electrons from Bhabha scattering. The EMC measures photon energies with a resolution of $2.5\%$ ($5\%$) at $1$~GeV in the barrel (end cap) region. The time resolution in the TOF barrel region is 68~ps, while that in the end cap region is 110~ps. The end cap TOF system was upgraded in 2015 using multigap resistive plate chamber technology, providing a time resolution of 60~ps~\cite{etof1, etof2, etof3}.

\section{Results from BESIII}

\subsection{Helicity amplitude analysis of $\chi_{cJ} \to \phi \phi$~\cite{BESIII:2023zcs}}

In the quark model, the $\chi_{cJ}$ states are identified as $P$-wave triple charmonium states with spin, parity and charge conjugation $J^{++}~(J=0,1,2)$. At leading order, the hadronic decays of $\chi_{cJ}$ are described by annihilations of charm and anti-charm quarks into two gluons and subsequent production of light and/or strange quarks. Early theoretical calculations for exclusive decays of $\chi_{cJ}$ into light hadrons have yielded smaller branching fractions than experimental measurements~\cite{Duncan:1980qd, Jones:1981ff, Anselmino:1992rw}.

Following perturbative QCD calculations~\cite{Zhou:2004mw}, the $\chi_{c1}$ decay rate should be strongly suppressed compared to $\chi_{c0}$ and $\chi_{c2}$, due to the helicity selection rule~\cite{Chernyak:1981zz} and the requirement of identical particle symmetry~\cite{Yang:1950rg}. However, the previous BESIII measurement reported similar branching fractions of $\chi_{cJ} \to \phi \phi$ decays for $\chi_{c0}$, $\chi_{c1}$ and $\chi_{c2}$~\cite{BESII:2011hcd}, namely $\mathcal{B}(\chi_{c0} \to \phi \phi) = (7.8 \pm 0.4 \pm 0.8) \times 10^{-4}$, $\mathcal{B}(\chi_{c1} \to \phi \phi) = (4.1 \pm 0.3 \pm 0.4) \times 10^{-4}$ and $\mathcal{B}(\chi_{c2} \to \phi \phi) = (10.7 \pm 0.4 \pm 1.1) \times 10^{-4}$. Meanwhile, the analysis of $\phi$ polarization will also be a key measurement to probe hadronic-loop effects in the $\chi_{cJ} \to \phi \phi$ decays~\cite{Huang:2021kfm}. Moreover, the ratios of the helicity amplitudes are found to be effective in the discrimination between the proposed models as these ratios are less sensitive to the parameters used in the evaluation of the model prediction~\cite{Zhou:2004mw, Chen:2013gka, Huang:2021kfm}. Table~\ref{tab:ratio_amplitudes} summarizes the helicity-amplitude ratios predicted by the considered theoretical models, where the uncertainties are due to the uncertainties on parameters involved in the calculation. The variable $x$ is defined as the ratio of transverse over the longitudinal polarized helicity amplitudes of the $\phi$ meson in $\chi_{c0}\to\phi\phi$: $x = \left|F^{0}_{1,1}/F^{0}_{0,0}\right|$ and the variables $\omega_{i}$ $(i=1,2,4)$ indicate the ratios of transverse over longitudinal polarized helicity amplitudes of the $\phi$ meson in $\chi_{c2}\to\phi\phi$: $\omega_1 = \left| F^{2}_{0,1}/F^{2}_{0,0}\right|$, $\omega_2 = \left| F^{2}_{1,-1}/F^{2}_{0,0}\right|$, $\omega_4 = \left| F^{2}_{1,1}/F^{2}_{0,0}\right|$, where $\lambda_1$ and $\lambda_2$ are the helicities of the two $\phi$, and $F^{J=0,2}_{\lambda_1,\lambda_2}$ are the helicity amplitudes. The $\chi_{c1}\to\phi\phi$ helicity amplitudes allow to test the validity of the identical particle symmetry: in this context the helicity-amplitude ratios  $u_1=|F^1_{1,0}/F^1_{0,1}|$ and $u_2=|F^1_{1,1}/F^1_{1,0}|$ are expected to be 1 and 0, respectively~\cite{Chen:2013gka}.

\begin{table}[htbp]
    \caption{Numerical predictions of the helicity-amplitude ratios from pQCD~\cite{Zhou:2004mw}, $^3P_0$~\cite{Chen:2013gka} and $D\bar D$ loop models ~\cite{Huang:2021kfm}.}
    \centering
    \begin{tabular}{l|c|ccc}
    \hline
    \hline
                Decay channel     &$\chi_{c0} \to \phi\phi$     &\multicolumn{3}{c}{$\chi_{c2} \to \phi\phi$}\\\hline
                Parameter         &$x$                        &$\omega_1$  &$\omega_2$  &$\omega_4$\\
		\hline
                pQCD              &$0.293 \pm 0.030$            &$0.812 \pm 0.018$     &$1.647 \pm 0.067$      &$0.344 \pm 0.020$\\
                $^3P_0$           &$0.515 \pm 0.029$            &$1.399 \pm 0.580$     &$0.971 \pm 0.275$      &$0.406 \pm 0.017$\\
        $D\bar{D}$ loop   &$0.359\pm0.019$  &$1.285\pm0.017$ &$5.110\pm0.057$ &$0.465\pm0.002$\\
    \hline
	    \hline
    \end{tabular}
    \label{tab:ratio_amplitudes}
\end{table}

\begin{figure}[htbp]
	\centering
	\includegraphics[width=0.8\textwidth]{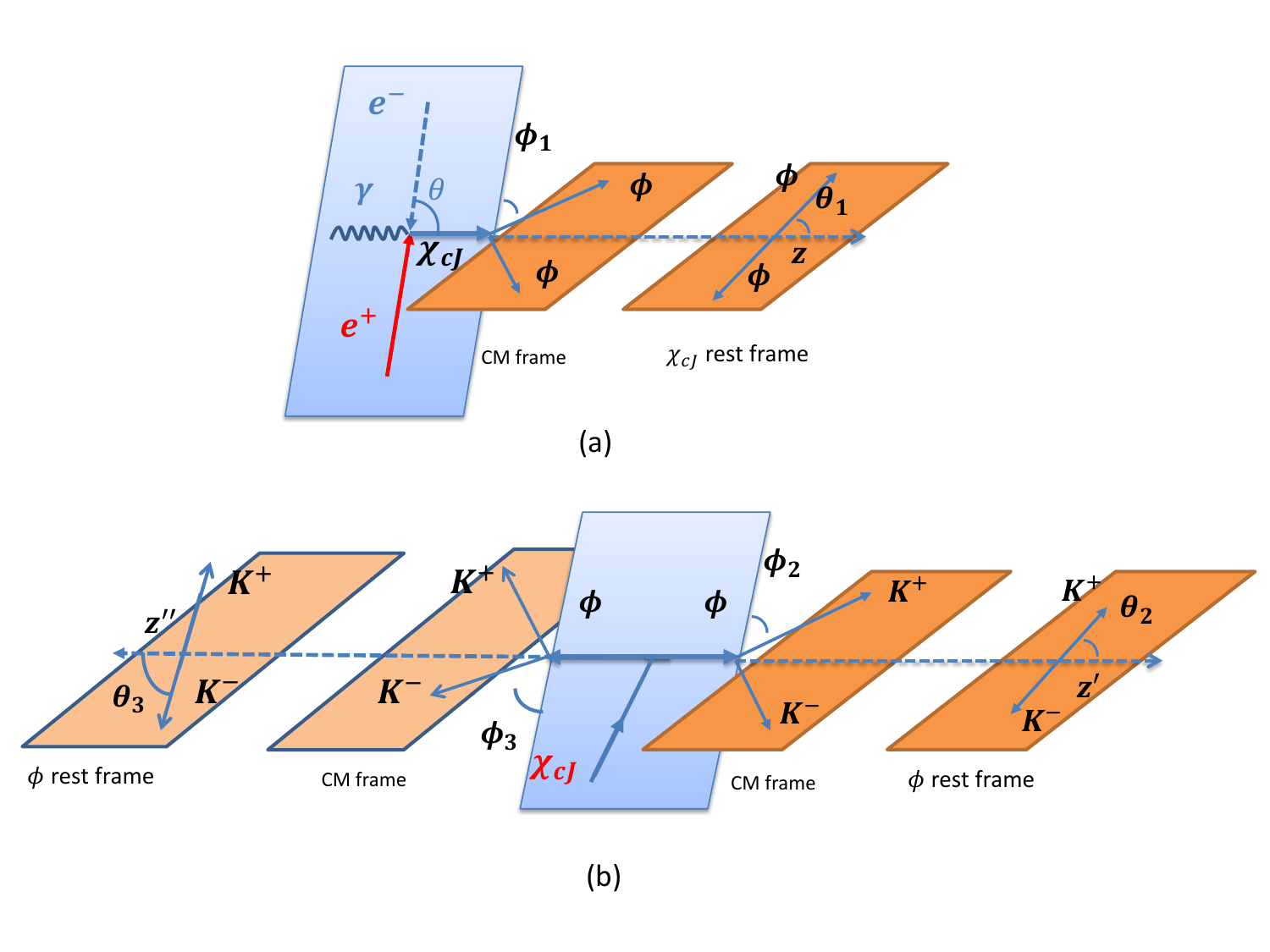}
	\caption{Definitions of helicity angles.}
	\label{fig:helicity_system}
\end{figure}

The amplitude analysis of $\chi_{cJ} \to \phi \phi$ is performed based on $448.1 \times 10^{6}$ $\psi(3686)$ events. The joint distribution for the sequential decays $e^+ e^- \to \psi(3686) \to \gamma \chi_{cJ},~\chi_{cJ} \to \phi \phi$ and $\phi \to K^+ K^-$ is constructed in the helicity system as shown in Figure~\ref{fig:helicity_system}. The joint amplitude for the sequential process is described as Eq.~(\ref{ampform}) and (\ref{BWform}).

\begin{eqnarray}\label{ampform}
\mathcal{M}(R_i)  &=& {1\over 2}\sum_{M,\lambda_{R},\lambda_{1},\lambda_{2}}
 A_{\lambda_{R},\lambda_{\gamma}}^{1}D_{M,\lambda_{R}-\lambda_{\gamma}}^{1*}(0,\theta_{0},0) F_{\lambda_{1},\lambda_{2}}^{J}D_{\lambda_{R},\lambda_{1}-\lambda_{2}}^{J*}(\phi_{1},\theta_{0},0)\nonumber\\
 &\times&B_{0,0}^{1}D_{\lambda_{1},0}^{1*}(\phi_{2},\theta_{2},0)B_{0,0}^{1}D_{\lambda_{2},0}^{1*}(\phi_{3},\theta_{3},0)BW(m_{\phi\phi},m_{i},\Gamma_{i}),
\end{eqnarray}
with
\begin{eqnarray}\label{BWform}
BW(m_{\phi\phi},m_i, \Gamma_i) = {1 \over m_{\phi\phi}^2-m_i^2 + i m_i \Gamma_i }.
\end{eqnarray}

The partial decay rate of $\psi(3686)$ is given by
\begin{equation}
\mathrm{d}\sigma \propto \frac{1}{2}{\sum}_{M,\lambda_{\gamma}}\left|\sum_{R_i}\mathcal{M}(R_i)\right|^2\mathrm{d}\Phi,
\end{equation}
where $\mathrm{d}\Phi$ is the standard phase space for the decay $\psi(3686)\to\gamma\phi\phi$ with $\phi\to K^+ K^-$. The detailed definitions of the symbols in the above equations can be found in the published paper~\cite{BESIII:2023zcs}.

The potential interference between the $\chi_{c0}$ and non-resonant contributions is considered while the interference of $\chi_{c1}$ and $\chi_{c2}$ with the non-resonant constribution is neglected due to the quite narrow width.

\begin{figure}[htbp]
\begin{center}
\mbox{
   \begin{overpic}[scale=0.35]{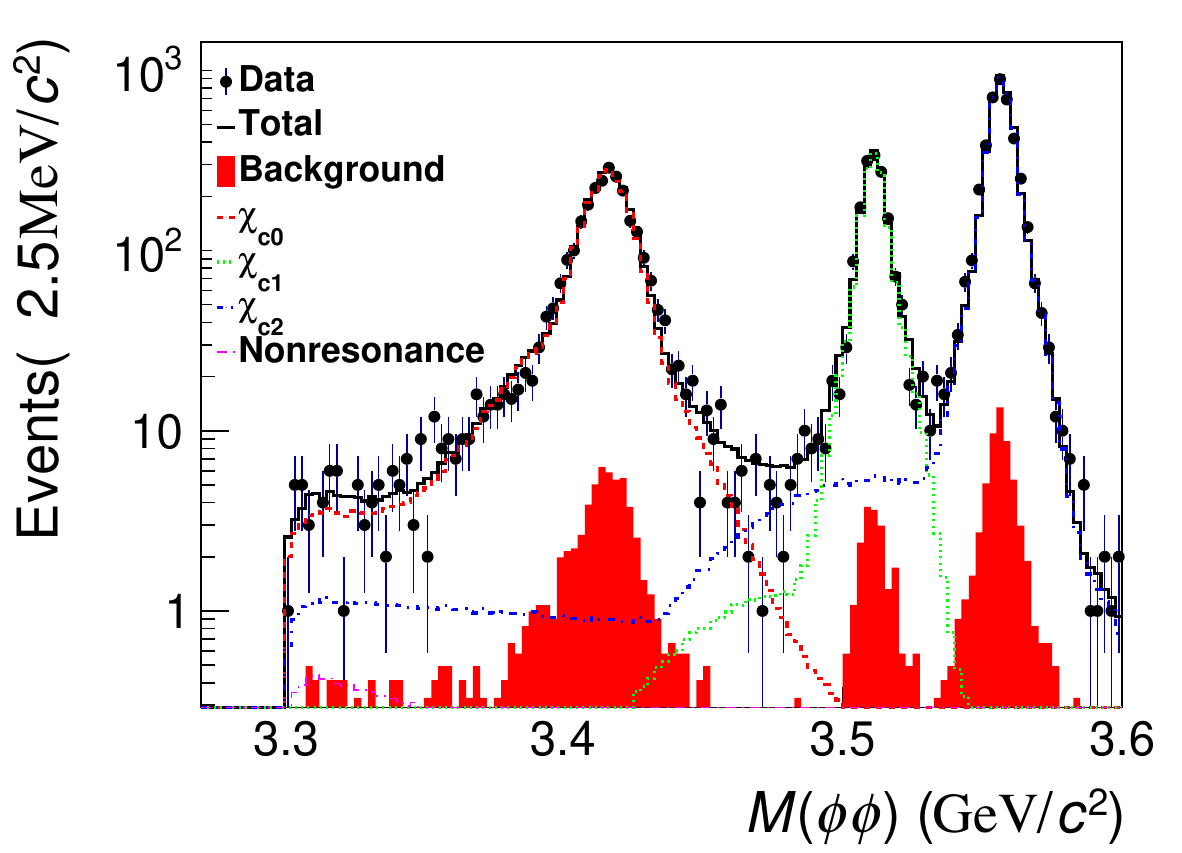}\end{overpic}
   \begin{overpic}[scale=0.35]{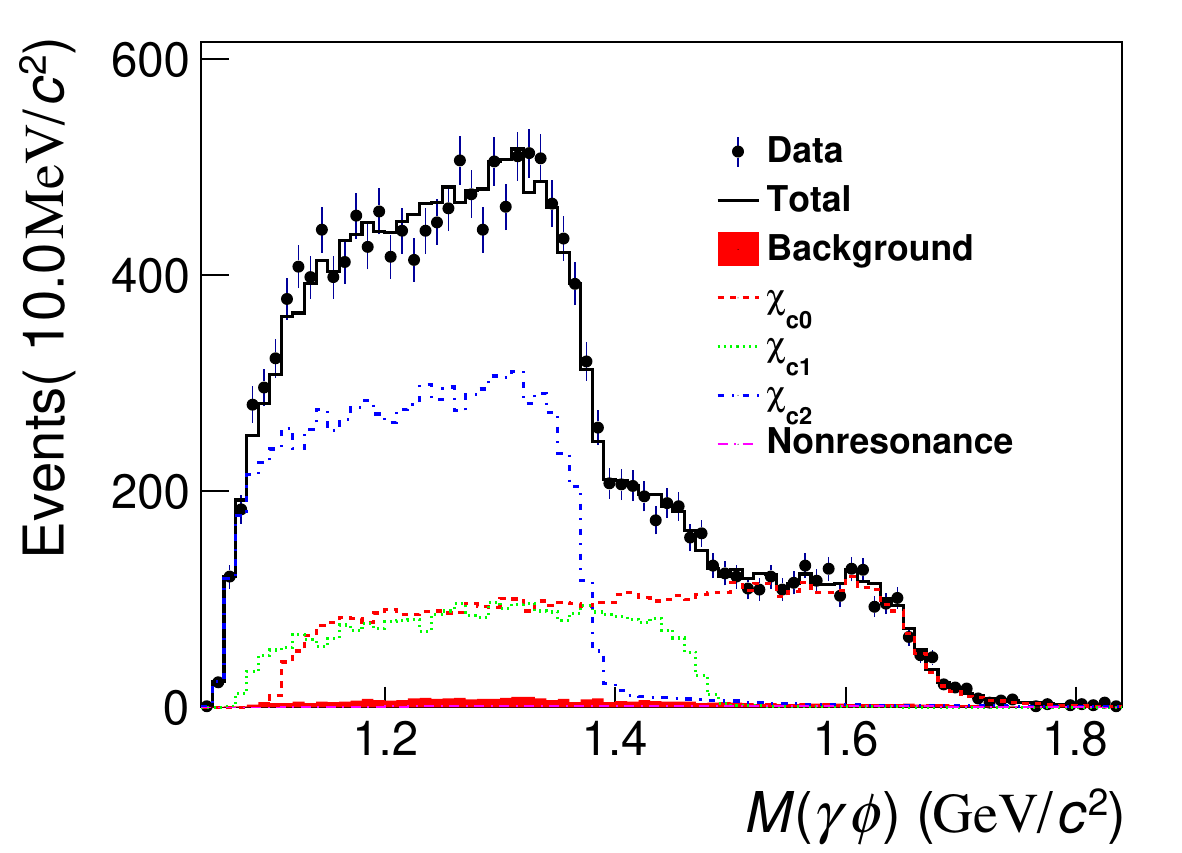}\end{overpic}
   }
\caption{Fit results of invariant mass distributions, $m_{\phi\phi}$ in the log version (left) and $m_{\gamma\phi}$ (right). The points with error bars represent data events. The black solid curve denotes the total fit result. The $m_{\gamma\phi}$ distribution has
two entries per event. Distributions of non-resonant events are almost invisible owing to the small contribution of this component.
   \label{fig:fitchicj}}
\end{center}
\end{figure}

The fit result is shown in Figure~\ref{fig:fitchicj}.

The branching fractions for $\chi_{cJ} \to \phi \phi$ are measured to be
\begin{eqnarray}
\mathcal{B}(\chi_{c0} \rightarrow \phi \phi)&=& (8.48\pm0.26\pm0.27)\times10^{-4} ,\nonumber \\
\mathcal{B}(\chi_{c1} \rightarrow \phi \phi)&=& (4.36\pm0.13\pm0.18)\times10^{-4} , \\
\mathcal{B}(\chi_{c2} \rightarrow \phi \phi)&=& (13.36\pm0.29\pm0.49)\times10^{-4}\nonumber,
\end{eqnarray}
where the first uncertainties are statistical and the second systematic. Comparing these results with BESIII previous measurement~\cite{BESII:2011hcd} and PDG values~\cite{ParticleDataGroup:2022pth}, as reflected in the Table \ref{tab BFsum}, the precision is improved by a factor of about $2$, but the values are greater.

\begin{table}[htbp]
\begin{center}
\caption{ \label{tab BFsum} Comparsion of measured branching fractions (BF).}
\begin{normalsize}
\begin{tabular}{ c|c|c|c }
\hline
\hline
Decay Mode & BF(2011 BESIII)~\cite{BESII:2011hcd} & BF(this work)& BF(PDG value)~\cite{ParticleDataGroup:2022pth}\\
\hline
Br[$\chi_{c0}$ $\rightarrow$ $\phi \phi$]($\times$$10^{-4}$) &  7.8$\pm$0.4$\pm$0.8 & $8.48\pm0.26\pm0.27$ & 7.7$\pm$0.7\\
Br[$\chi_{c1}$ $\rightarrow$ $\phi \phi$]($\times$$10^{-4}$) &  4.1$\pm$0.3$\pm$0.5 & $4.36\pm0.13\pm0.18$ & 4.2$\pm$0.5\\
Br[$\chi_{c1}$ $\rightarrow$ $\phi \phi$]($\times$$10^{-4}$) &  10.7$\pm$0.4$\pm$1.2 & $13.36\pm0.29\pm0.49$ & 11.2$\pm$1.0\\
\hline\hline
\end{tabular}
\end{normalsize}
\end{center}
\end{table}

The ratios of the amplitude moduli are measured to be
\begin{eqnarray}
\left|F_{1,1}^0\right|/\left|F_{0,0}^0\right| &=& 0.299\pm0.003\pm0.019,
\end{eqnarray}
for $\chi_{c0}\to\phi \phi$, and
\begin{eqnarray}
\left|F_{0,1}^2\right|/\left|F_{0,0}^2\right| &=& 1.265\pm0.054\pm0.079,\\
\left|F_{1,-1}^2\right|/\left|F_{0,0}^2\right| &=& 1.450\pm0.097\pm0.104,\\
\left|F_{1,1}^2\right|/\left|F_{0,0}^2\right| &=& 0.808\pm0.051\pm0.009,
\end{eqnarray}
for $\chi_{c2}\to\phi \phi$, where the first and second uncertainties are statistical and systematic, respectively. Additionally, there is no evidence of identical particle symmetry breaking from the study of $\chi_{c1}\to\phi \phi$.

\begin{figure}[htbp]
\begin{center}
\mbox{
   \begin{overpic}[scale=0.6]{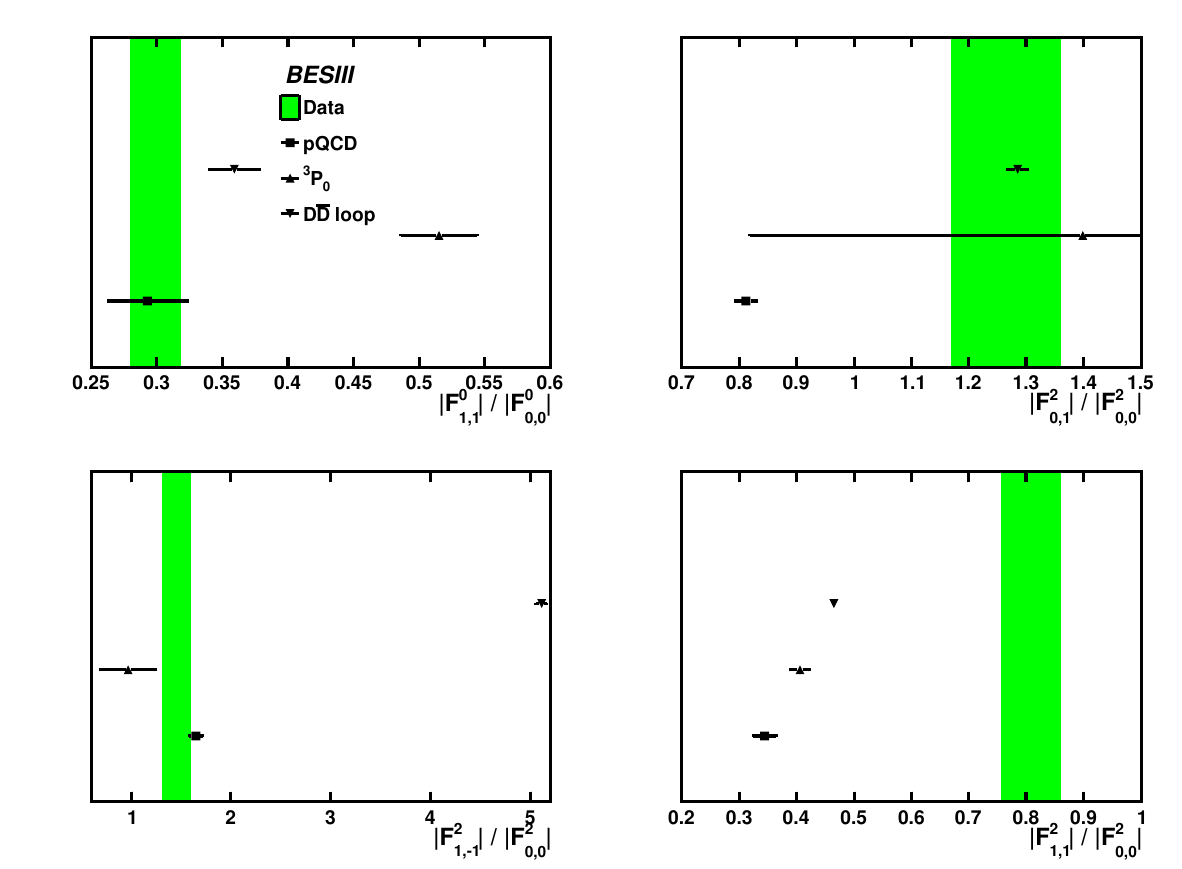} \end{overpic}
   }
\caption{Comparison of the measured amplitude ratios with the predicted ones from pQCD, the $^3P_0$ model and the $D\bar {D}$ loop model.
\label{cmptheo}}
\end{center}
\end{figure}

Figure~\ref{cmptheo} shows a comparison of the measured amplitude ratios to the corresponding theoretical predictions. The measured ratio of amplitude moduli for the $\chi_{c0}$ is consistent with the pQCD prediction of Ref.~\cite{Zhou:2004mw}, since two independent helicity amplitudes of the $\chi_{c0}\to\phi\phi$ decay,  $F_{1,1}^0$ and $F_{0,0}^0$,  follow the helicity selection rule. For the $\chi_{c2}$ decay, the measured ratios of amplitude moduli deviate from the pQCD~\cite{Zhou:2004mw}, $^3P_0$~\cite{Chen:2013gka} and $D\bar {D}$ loop~\cite{Huang:2021kfm} predictions with $\chi^2/\mathrm{ndf}=23.2$, $23.8$, and $155.2$, respectively. The $D\bar D$ loop model can be ruled out due to the large deviation. However, the predictions of other models also differ from the experimental results. In short, all of the above theories use some of the input from the experimental results, thus this measurement can provide more constraints for further developing the models. It could also be a basis for the measurement in the future, as 2.7 billion $\psi(3686)$ events have been accumulated in BESIII~\cite{BESIII:2020nme}.

\subsection{Evidence for the $\eta_c(2S) \to \pi^+ \pi^- \eta$ decay~\cite{BESIII:2022ksv}}

	Until now, the knowledge about $\eta_c(2S)$ is still limited~\cite{Brambilla:2010cs}, suffering from the very soft photon from $\psi(3686) \to \gamma \eta_c(2S)$. The total measured branching fraction of $\eta_c(2S)$ decays is small (less than 5\%) according to PDG~\cite{ParticleDataGroup:2022pth}.

	The decay of charmonium states into light hadrons is believed to be dominated by the annihilation of the $c\bar{c}$ pair into two or three gluons. The so-called ``12\% rule'' states that the ratio of the inclusive branching fractions of light hadron states between $\psi(3686)$ and $J/\psi$ is about 12\%~\cite{Appelquist:1974zd}. Similarly, one would expect a similar ratio of the hadronic branching fractions between $\eta_c(2S)$ and $\eta_c$ due to their analogous decaying dynamics in comparison to $\psi(3686)$ and $J/\psi$. According to Ref.~\cite{Franklin:1983ve}, for many normal light hadronic channel $h$,
\begin{equation}
	\frac{\mathcal{B}(\eta_c(2S) \to h)}{\mathcal{B}(\eta_c \to h)} \approx \frac{\mathcal{B}(\psi(3686) \to h)}{\mathcal{B}(J/\psi \to h)} = 0.128,
\end{equation}
	while there are also theoretical works resulting in a ratio close to one~\cite{Chao:1996}. The measured ratios are mostly greater than 12\% and less than one, except the ones with $p \bar{p}$ final states, so the further measurements of the decays of $\eta_c(2S)$ and $\eta_c$ are of great significance.

\begin{figure}[htbp]
	\centering
	\includegraphics[width=0.7\textwidth]{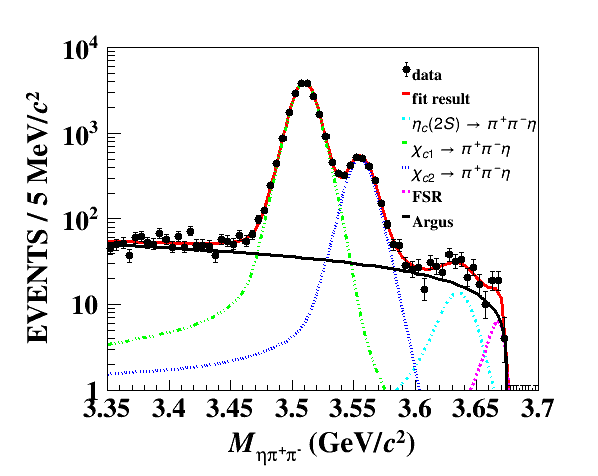}
	\caption{The result of a fit to the invariant mass distribution of ${\pi^+ \pi^- \eta}$. The black dots with error bars are BESIII data, the red and black solid curves denote the total fit curve and the shape of the smooth background contributions, respectively. The green dash-dot-dotted curve, the blue dotted, and the cyan dash-dotted denote the decay modes of $\chi_{c1}$, $\chi_{c2}$ and $\eta_c(2S)$, while the pink dashed curve denotes the contributions of the FSR process $\psi(3686)\to(\gamma_{\rm FSR})\pi^+\pi^-\eta$.}
	\label{fig:fit_etapipi}
\end{figure}

	Based on $448.1 \times 10^{6}$ $\psi(3686)$ events collected at BESIII, the decay $\eta_c(2S) \to \pi^+ \pi^- \eta$ is searched for. The final fit result is shown in Figure~\ref{fig:fit_etapipi}.

Evidence for the decay $\eta_c(2S)\to\pi^+\pi^-\eta$ is found for the first time, with a statistical significance of 3.5$\sigma$. The product of the branching fractions is measured to be $Br(\psi(3686)\to\gamma\eta_c(2S))\times Br(\eta_c(2S)\to\pi^+\pi^-\eta)=$  (2.97$\pm$0.81$\pm$0.26)$\times10^{-6}$, where the first uncertainty is statistical and the second systematic. The branching fraction of $\eta_c(2S)$ decaying into $\pi^+ \pi^- \eta$ is measured to be $Br(\eta_c(2S)\to\pi^+\pi^-\eta)=$  $(42.4\pm11.6\pm3.8\pm30.3)\times10^{-4}$, with the first uncertainty being the statistical, the second the systematic uncertainty without taking into account the uncertainty of the branching fraction of $\psi(3686)\to\gamma\eta_c(2S)$. The third one is the systematic uncertainty arising from this branching fraction.

\begin{figure}[htbp]
\begin{center}
\begin{overpic}[width=0.7\textwidth]{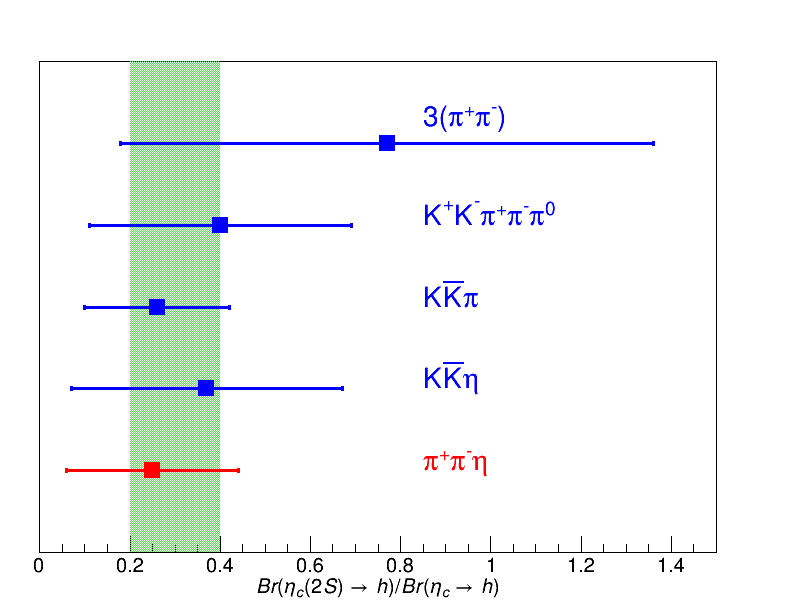}
\end{overpic}
\caption{Estimation of the averaged value of the ratio of Br($\eta_c(2S)\to h$) to Br($\eta_c\to h$). Here, $h$ means various hadronic final states, as shown on this figure. Except the branching fraction of $\eta_c(2S) \to \pi^+ \pi^- \eta$ of this work, the other results are quoted from Refs.~\cite{ParticleDataGroup:2022pth,BESIII:2022hcv}.  The shade is the averaged value of these five decay modes with one standard deviation. Both statistical and systematic uncertainties have been included.}
\label{fig::comp_br}
\end{center}
\end{figure}

With the branching fraction $Br(\eta_c\to\pi^+\pi^-\eta)=$(1.7$\pm$0.5)\%~\cite{ParticleDataGroup:2022pth}, the ratio of the branching fractions of $\eta_c$ and $\eta_c(2S)$ decaying into $\pi^+ \pi^- \eta$ is calculated to be $\frac{Br(\eta_c(2S)\to\pi^+\pi^-\eta)}{Br(\eta_c\to\pi^+\pi^-\eta)}=0.25\pm0.20$. Combining the ratios of other hadronic decay modes of $\eta_c(2S)$ to $\eta_c$~\cite{ParticleDataGroup:2022pth,BESIII:2022hcv}, the averaged value of all these ratios including this measurement is determined to be 0.30$\pm$0.10 (see Figure~\ref{fig::comp_br}). This ratio agrees neither with the prediction in Ref.~\cite{Franklin:1983ve} nor in Ref.~\cite{Chao:1996}. With about 2.7 billion $\psi(3686)$ events to be accumulated, BESIII will make a further substantial contribution to this field~\cite{BESIII:2020nme}.

\subsection{Observation of $\psi(3770) \to \eta J/\psi$~\cite{BESIII:2022yoo}}

Conventionally, the $\psi(3770)$  has been regarded as the lowest-mass $D$-wave charmonium state above the $D\bar{D}$ threshold, {\it i.e.} a pure $c\bar{c}$ meson in the quark model~\cite{Eichten:1978tg}, while this model cannot explain the measured large non-$D\bar{D}$ decay width of the state~\cite{He:2008xb, ParticleDataGroup:2022pth}. To solve this puzzle, various theoretical models are developed, either by introducing tetra-quark component into the wave function~\cite{Voloshin:2005sd}, or more complicated dynamics such as $2S-1D$ mixing between $\psi(3686)$ and $\psi(3770)$~\cite{Mo:2006cy, Rosner:2004wy, Zhang:2009gy, Ding:1991vu}, and re-scattering mechanism with $D$ mesons~\cite{Guo:2012tj, Wang:2011yh, Liu:2009dr, Guo:2010ak}. Until now, the only well-established non-$D\bar{D}$ channel is $\psi(3770) \to \pi^+ \pi^- J/\psi$~\cite{ParticleDataGroup:2022pth, Eichten:2007qx} so that the further sutdy of other non-$D\bar{D}$ channels is necessary and essential in both experimental and theoretical aspects. In 2005, CLEO studied the decay $\psi(3770) \to \eta J/\psi$ and reported the branching fraction to be $(8.7 \pm 3.3 \pm 2.2) \times 10^{-4}$ at a statistical significance of $3.5\sigma$ without considering the interference~\cite{CLEO:2005zky}. The branching fraction of $\psi(3770) \to \eta J/\psi$ is utilized as an input in theoretical calculations of decay properties not only for conventional charmonium states~\cite{Zhang:2009kr, Li:2013zcr} but also for exotic charmonium-like (also called $XYZ$) states~\cite{Anwar:2016mxo} observed in this energy region.

\begin{figure}[hbtp]
\centering
\subfigure{
\begin{minipage}{0.3\textwidth}
\centering
\includegraphics[width=\textwidth]{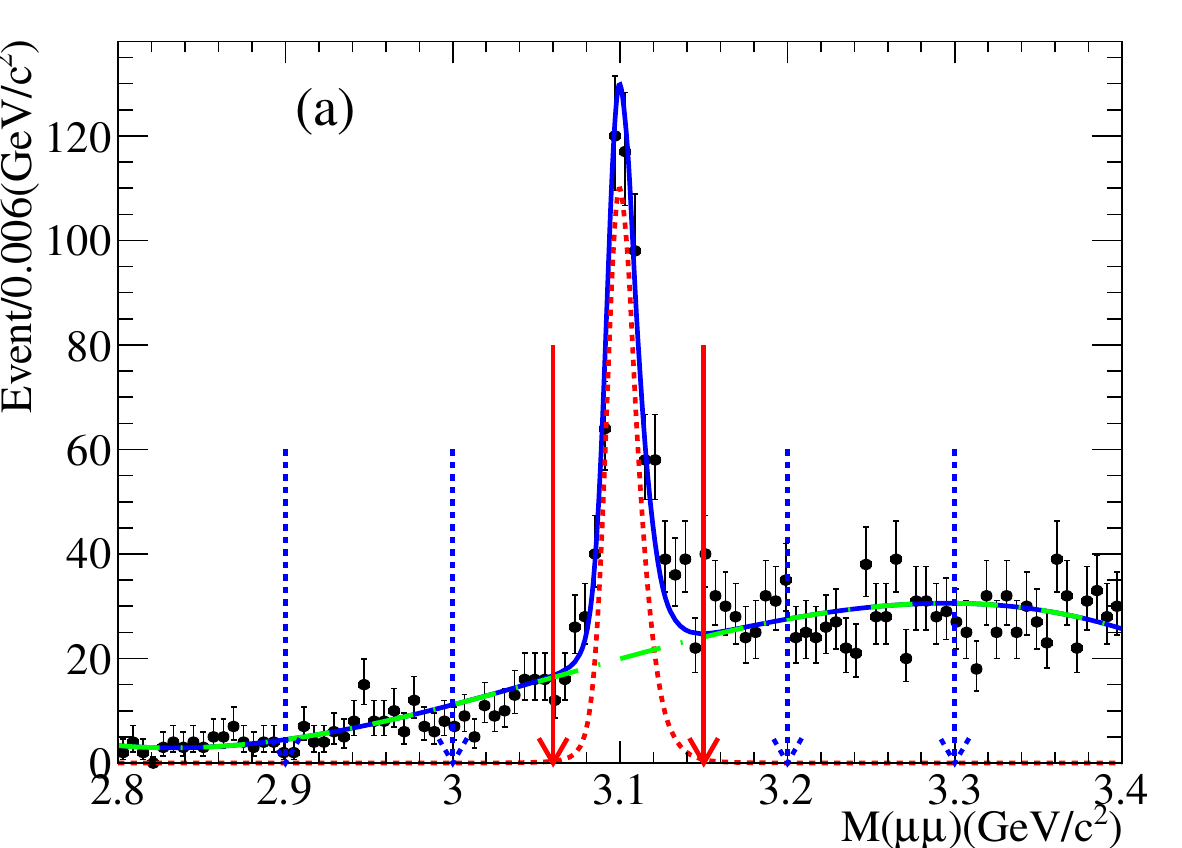}
\end{minipage}
\begin{minipage}{0.3\textwidth}
\centering
\includegraphics[width=\textwidth]{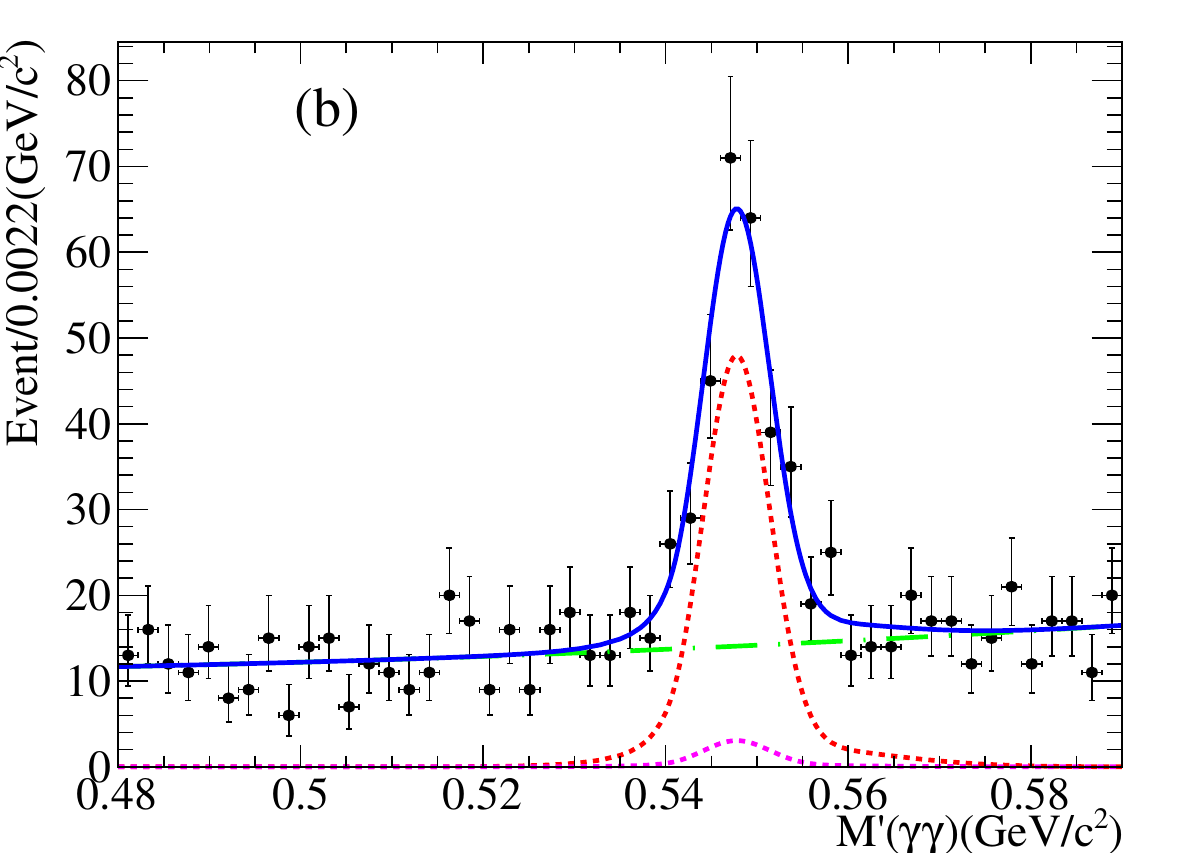}
\end{minipage}
\begin{minipage}{0.3\textwidth}
\centering
\includegraphics[width=\textwidth]{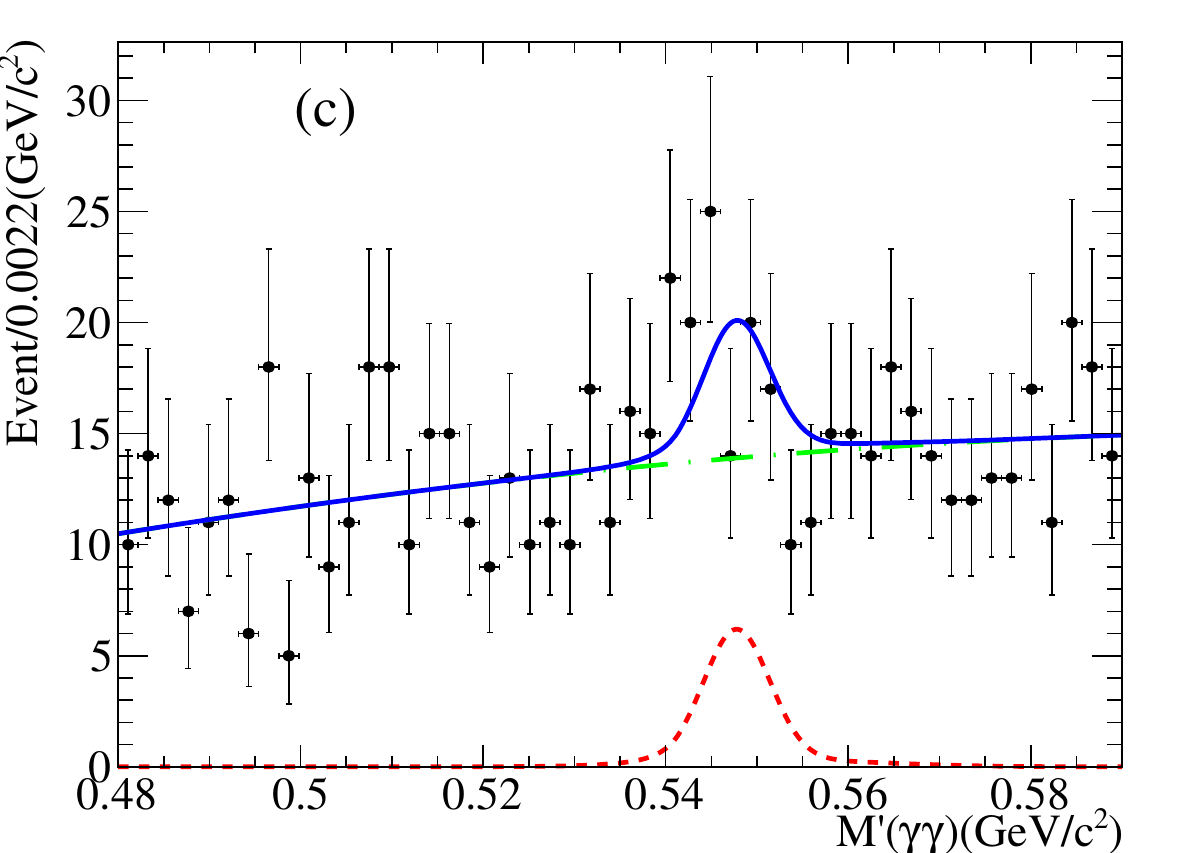}
\end{minipage}}
\caption{ Distribution of $M(\mu\mu)$ from data (a). The signal region is indicated by the two solid red arrows, while the sideband regions by the two dashed blue arrows. Distributions of $M'(\gamma \gamma)$ in the $J/\psi$ signal region (b) and sideband regions (c). The points with error bars are data, the blue solid curves represent the fit results, the red dashed curves represent signal components, and the green dot-dashed curves represent background components.}
\label{fig:sig_yield_eta_jpsi}
\end{figure}

The signal yield $N_{\rm obs}$ is obtained by fitting to the $M^{\prime}(\gamma \gamma)$ ($M^{\prime}(\gamma \gamma) \equiv M(\gamma \gamma) + M(\mu \mu) - m_{J/\psi}$) distribution as shown in Figure~\ref{fig:sig_yield_eta_jpsi}. Then the Born cross section is calculated by
\begin{equation}
{\sigma^{B}(e^+e^-\rightarrow\eta J/\psi)}= \frac{N^ {\rm obs}}{\mathcal{L}\cdot(1+\delta^{\rm ISR})\cdot(1+\delta^{\rm VP})\cdot\varepsilon\cdot{\cal{B}}r} \,,
\label{eq:born}
\end{equation}
where $\mathcal{L}$ is the integrated luminosity, $(1+\delta^ {\rm ISR})$ is the ISR correction factor~\cite{Kuraev:1985hb}, $(1+\delta^{\rm VP})$ is the vacuum polarization factor taken from QED calculation~\cite{WorkingGrouponRadiativeCorrections:2010bjp}, ${\cal{B}}r$ is the product of the branching fractions of the subsequent decays of intermediate states as given by the PDG~\cite{ParticleDataGroup:2022pth}, and $\varepsilon$ is the detection efficiency. The ISR correction factor is obtained by an iterative method~\cite{Sun:2020ehv}, in which the dressed cross section measured in this study and previously with c.m.\ energies from $\sqrt{s}=3.81$ to $4.60~{\rm GeV}$~\cite{BESIII:2020bgb} are used as input. Table~\ref{table5} shows the measured Born cross section at $\sqrt{s}=3.773~{\rm GeV}$ and the values of the other parameters in Eq.~\ref{eq:born}. Finally the Born cross section is calculated to be $\sigma^{B}(e^+ e^- \to \eta J/\psi) = (8.88 \pm 0.87_{\rm stat.} \pm 0.42_{\rm syst.})~{\rm pb}$ at $3.773~{\rm GeV}$.

\begin{table}[htbp]
	\centering
	\footnotesize
	\caption{ The values of the integrated luminosity $\mathcal{L}$, the signal yield $N_{\rm obs}$, the detection efficiency $\varepsilon$, the product of radiative correction factor and vacuum polarization factor $R = (1+\delta^{{\rm ISR}})\cdot(1+\delta^{VP})$, and the obtained Born cross section of $e^+e^-\rightarrow\eta J/\psi$ at $\sqrt{s}=3.773~{\rm GeV}$. The uncertainties on the efficiency and cross section are statistical only.}
\begin{tabular}{ccccccc}\hline\hline
	\multicolumn{1}{c}{\multirow {1}{*}{ $\mathcal L$ (pb$^{-1}$) }}  & \multicolumn{1}{c}{ $\varepsilon(\%) $ }  & \multicolumn{1}{c}{ $R$ }
	& \multicolumn{1}{c}{ ${\cal{B}}(J/ \psi\rightarrow\mu^+\mu^-)(\%)$ } & \multicolumn{1}{c}{ ${\cal{B}}(\eta\rightarrow\gamma\gamma)(\%)$ } &  \multicolumn{1}{c}{ $N^{{\rm obs}}$ } & \multicolumn{1}{c}{ $\sigma^B$(pb) } \\
\hline
 $2931\pm15$ &  $45.4\pm0.1$  & $0.80$ & $5.96\pm0.03$  & $39.4\pm0.2$ & $222\pm22$  &  $8.89\pm0.88$ \\\hline\hline
\end{tabular}
\label{table5}
\end{table}

The branching fraction of $\psi(3770) \to \eta J/\psi$ is obtained by fitting to the dressed cross section of $e^+ e^- \to \eta J/\psi$ from $\sqrt{s} = 3.773~{\rm GeV}$ to $4.60~{\rm GeV}$, combining the cross section in this work and the previous BESIII analysis~\cite{BESIII:2020bgb} with c.m.\ energies from $\sqrt{s}=3.81$ to $4.60~{\rm GeV}$. Two treatments of the $\psi(3770)$ resonant decay amplitude are considered. Assuming $\psi(3770)$ is coherent with the other amplitudes, we get
\begin{eqnarray}
	\sigma_{\rm co.} & =|C\cdot\sqrt{\Phi(s)}+e^{i\phi_1}{\rm BW}_{\psi(3770)}+e^{i\phi_2}{\rm BW}_{\psi(4040)} \nonumber  \\
			      & +e^{i\phi_3}{\rm BW}_{Y(4230)}+e^{i\phi_4}{\rm BW}_{Y(4390)}|^2.
\end{eqnarray}
If $\psi(3770)$ is incoherent with the other amplitudes, we will have
\begin{eqnarray}
\sigma_{\rm inco.} & = |{\rm BW}_{\psi(3770)}|^2 +|C\cdot\sqrt{\Phi(s)}+e^{i\phi_2}{\rm BW}_{\psi(4040)} \nonumber \\
                                & +e^{i\phi_3}{\rm BW}_{Y(4230)}+e^{i\phi_4}{\rm BW}_{Y(4390)}|^2 \, ,
\end{eqnarray}
where $\Phi(s)=q^{3}/s$ is the $P$-wave phase space factor used to parameterize the continuum term, with $q$ being the $\eta$ momentum in the $e^+ e^-$ c.m. frame, BW is the Breit-Wigner function, $\phi$ is the relative phase between the resonant decay and the phase space term, and $C$ is a real parameter.

\begin{figure}
\begin{minipage}{0.5\textwidth}
\centering
\begin{overpic}
[width=7cm] {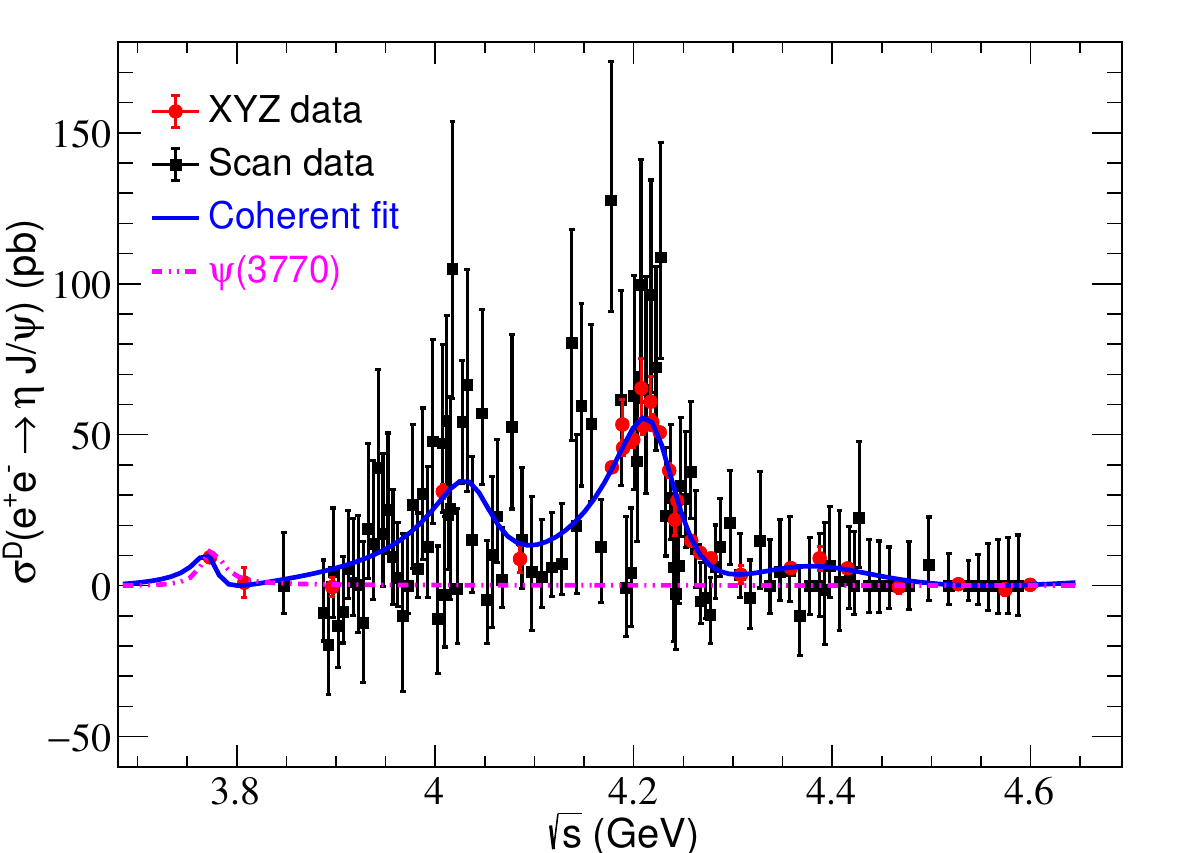}
\put(59, 40)
{\includegraphics[scale=0.12]{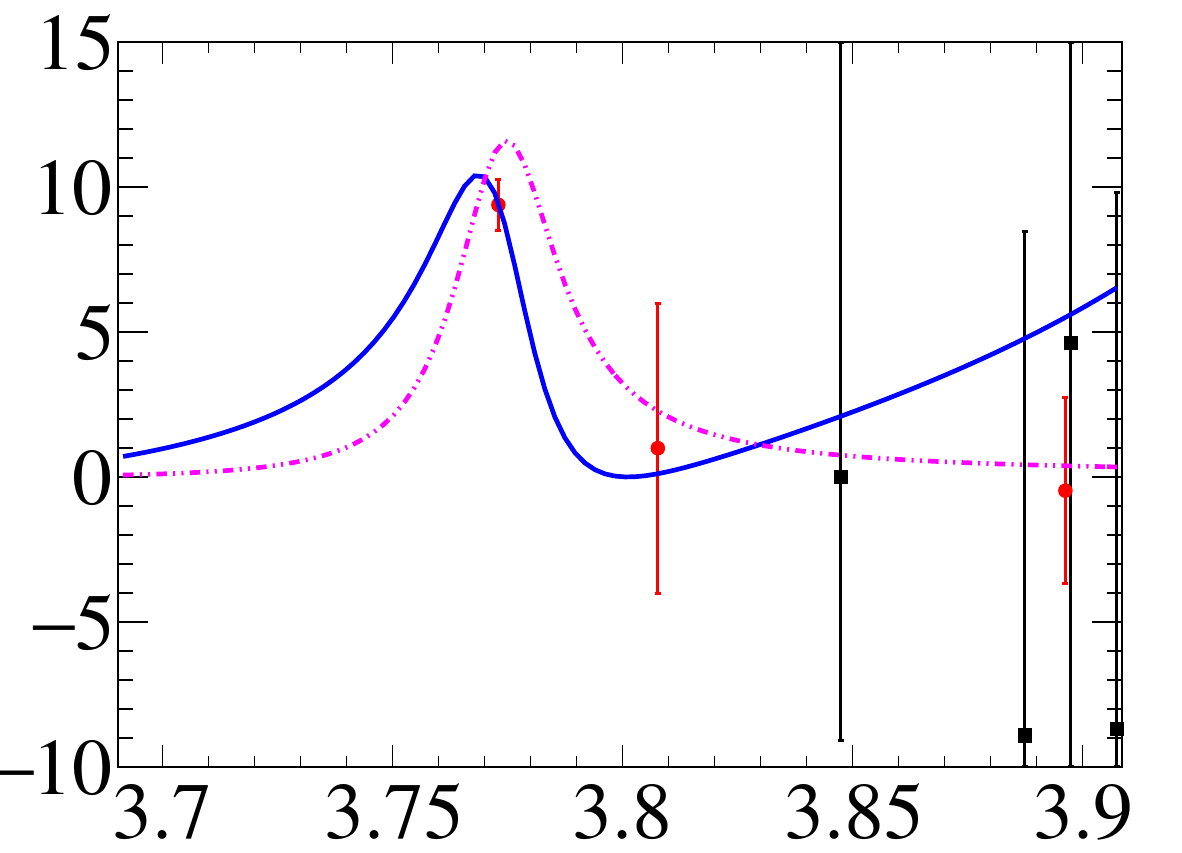}}
\end{overpic}
\end{minipage}
\begin{minipage}{0.5\textwidth}
\centering
\begin{overpic}
 [width=7cm]{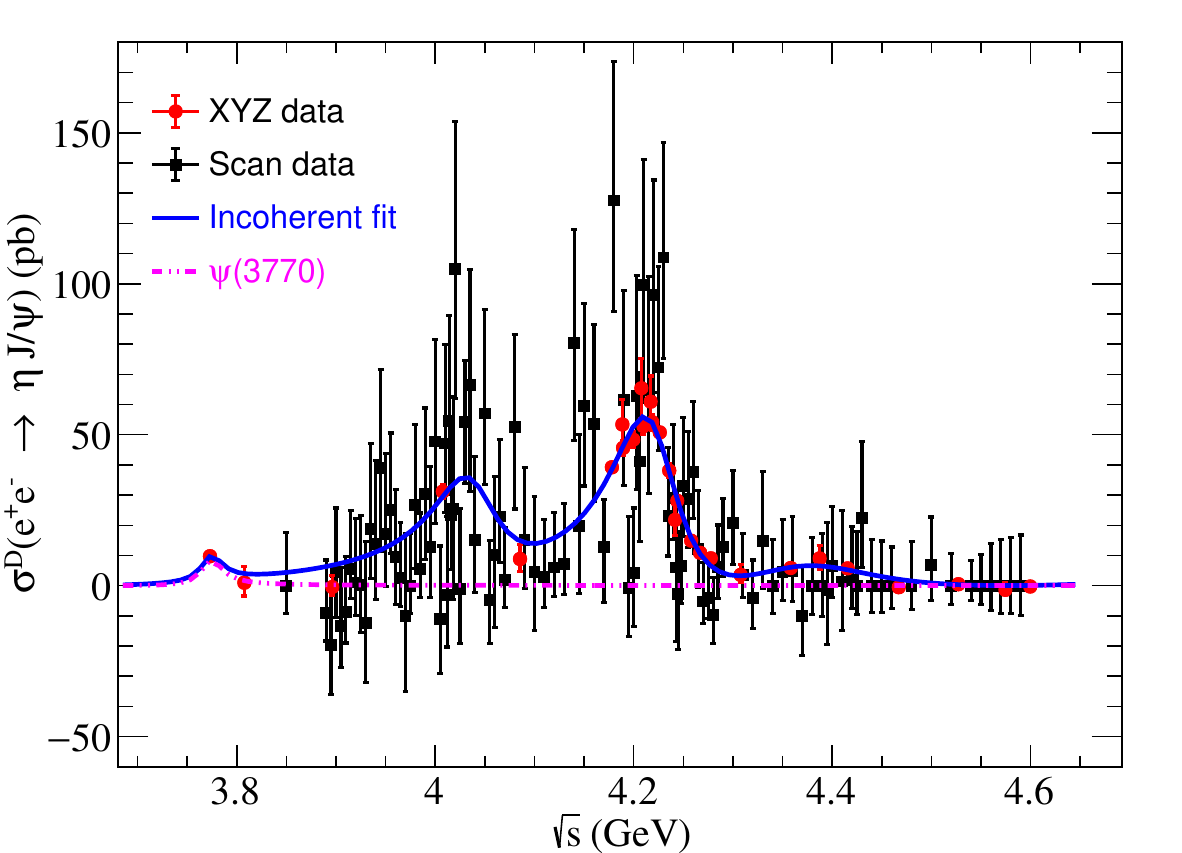}
 \put(59, 40)
{\includegraphics[scale=0.12]{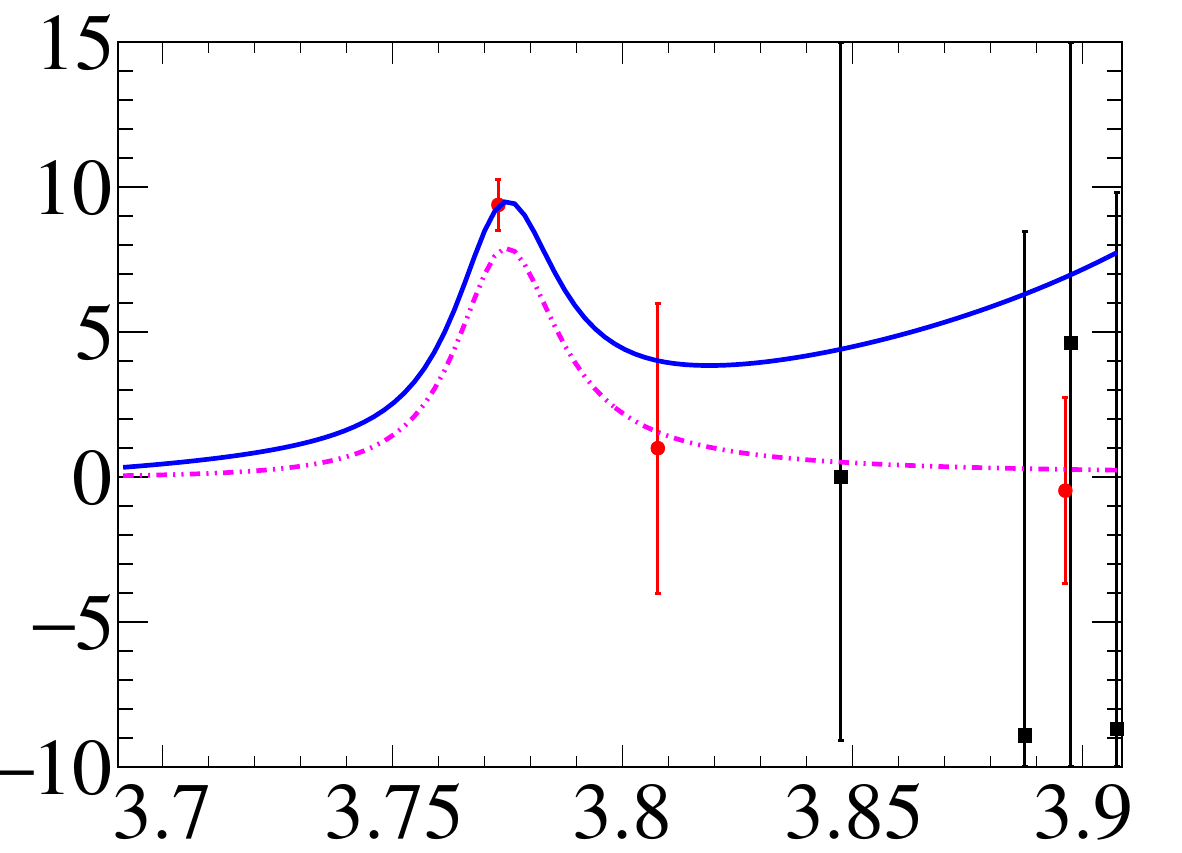}}
 \end{overpic}
\end{minipage}
\caption{(Left) Coherent and (right) incoherent fits to the dressed cross section line-shape of $e^+ e^- \to\eta J/\psi$. The points with error bars are data and the solid curves are the best fit results. The insert is the zoomed distribution in the $\psi(3770)$ mass region.}
\label{fig:fit_cross_section}
\end{figure}

Using the above two formulae, the dressed cross sections are fitted as shown in Figure~\ref{fig:fit_cross_section}. In the incoherent case, the branching fraction is determined to be $(8.7 \pm 1.0_{\rm stat.} \pm 0.8_{\rm syst.}) \times 10^{-4}$, close to the CLEO result~\cite{CLEO:2005zky}. In the coherent case, four solutions are obtained with branching fractions varying between $(11.6 \pm 6.1_{\rm stat.} \pm 1.0_{\rm syst.})\times10^{-4}$ and $(12.0 \pm 6.1_{\rm stat.} \pm 1.1_{\rm syst.})\times10^{-4}$. We suppose that there exists substantial interference effect, especially between $\psi(3770)$ and highly excited vector charmonium(-like) states.

\subsection{Observation of the decay $\chi_{cJ} \to \Omega^- \bar{\Omega}^+$~\cite{BESIII:2023bnk}}

The study of charmonium decays into baryon antibaryon ($B\bar{B}$) pairs provides a powerful tool for investigating many topics in quantum chromodynamics. In contrast to $J/\psi$ decays, the decays of the $P$-wave charmonium states, $\chi_{cJ}~(J=0,1,2)$, to $B\bar{B}$ have a non-trivial color-octet contribution~\cite{DASP:1975xwv, Feldman:1975bq}. Multiple models have been raised to describe $\chi_{cJ}$ to $B\bar{B}$ decays, including $p\bar{p}$, $\Lambda \bar{\Lambda}$, $\Sigma^+ \bar{\Sigma}^-$, $\Sigma^0 \bar{\Sigma}^0$, while none of them can describe all the final states~\cite{Wong:1999hc, Liu:2009vv}. Except the ground-state octet baryons above, it is desirable to extend these studies to decays of $\chi_{cJ}$ into pairs of decuplet ground-state baryons with spin 3/2. So far only $\chi_{c0} \to \Sigma(1385)^{\pm} \bar{\Sigma}(1385)^{\mp}$ decays~\cite{BESIII:2012wgp} have been studied by the BESIII Collaboration. The decay $\chi_{cJ} \to \Omega^- \bar{\Omega}^+$ is unique due to the presence of three pairs of strange quarks in the final state. This may give a distinct way for understanding quantum chromodynamics.

\begin{figure}[htbp]
        \begin{center}
    \includegraphics[width = 0.7\textwidth]{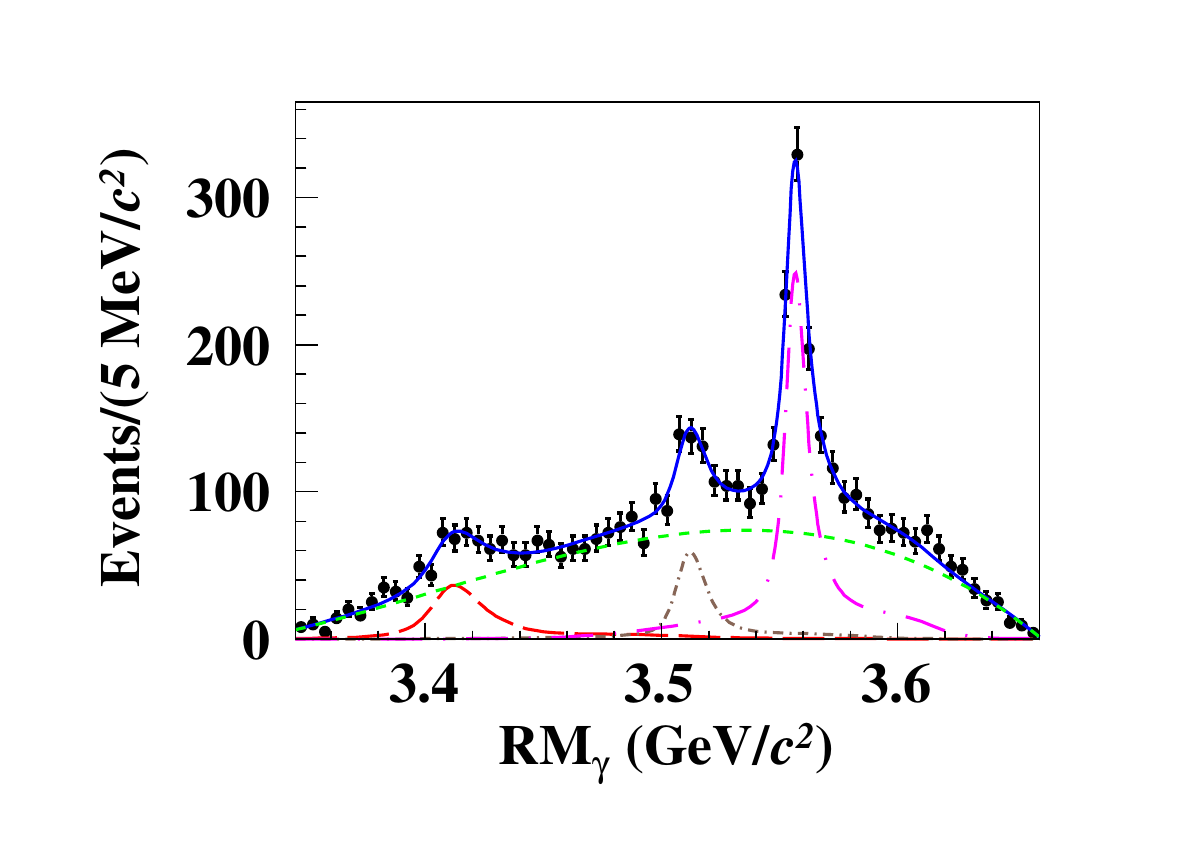}
        \end{center}
\vspace{-0.8cm}
        \caption{
                        Fit to the $RM_{\gamma}$ distribution in data.
            The dots with error bars are data, the blue solid line is the total fit,
                the green short dashed line represents the fitted combinatorial background shape, and
                the red long dashed, dark brown short dot-dashed and magenta long dot-dashed lines indicate the $\chi_{c0}$, $\chi_{c1}$ and $\chi_{c2}$ signals, respectively.
        }
        \label{fig:fit_RMgamma}
\end{figure}

Based on $2.708 \times 10^9$ $\psi(3686)$ events, the decay $\chi_{cJ} \to \Omega^- \bar{\Omega}^+$ is studied using the radiation decay $\psi(3686) \to \gamma \chi_{cJ}$. Signal yield is obtained by fitting to the recoil mass spectrum of the radiative photon ($RM_{\gamma}$) after partially reconstructing $\Omega^-(\bar{\Omega}^+)$. The fit result is shown in Figure~\ref{fig:fit_RMgamma}. Then the branching fraction is calculate by
\begin{equation}
    \mathcal{B}(\chi_{cJ} \to \Omega^-\bar{\Omega}^+) = \frac{N_{\chi_{cJ}}^{\rm obs}} {N_{\psi(3686)} \cdot \mathcal{B}_{\psi(3686) \to \gamma \chi_{cJ}} \cdot \epsilon_{\chi_{cJ}}},
\end{equation}
where $N_{\chi_{cJ}}^{\rm obs}$ is the signal yield, $N_{\psi(3686)}$ is the total number of $\psi(3686)$ events, $\epsilon_{\chi_{cJ}}$ is the detection
efficiency including the subsequent $\Omega$ and $\Lambda$ decays, and $\mathcal{B}_{\psi(3686) \to \gamma \chi_{cJ}}$ is the BF of the $\psi(3686) \to \gamma \chi_{cJ}$
decay~\cite{ParticleDataGroup:2022pth}. The measured branching fractions for the three signal modes are listed in Table~\ref{tab:yields}.

This is the first observation of $\chi_{cJ}$ decays into a pair of decuplet ground-state baryons with spin 3/2. The $\chi_{cJ} \to \Omega^- \bar{\Omega}^+$ decays can also be used to probe the spin polarization of $\Omega^-$ baryon in the charmonium production at the future tau-charm factories~\cite{Achasov:2023gey}.

\begin{table}[htbp]
    \begin{center}
    \caption{The $\chi_{cJ}$ signal yields ($N_{\chi_{cJ}}^{\rm obs}$), detection efficiencies ($\epsilon_{\chi_{cJ}}$), BFs of $\chi_{cJ} \to \Omega^- \bar{\Omega}^+$
    ($\mathcal{B}$) and the signal significances (${\rm Sig.}$). Here the uncertainties are statistical only.
    }
    \label{tab:yields}
    \setlength{\extrarowheight}{1.0ex}
    \renewcommand{\arraystretch}{1.0}
    \vspace{0.2cm}
 \begin{tabular}{p{1cm} | m{1.6cm}<{\centering} m{1.5cm}<{\centering} m{1.2cm}<{\centering} m{3.6cm}<{\centering}}
            \hline \hline
            Mode & $N_{\chi_{cJ}}^{\rm obs}$ & $\epsilon_{\chi_{cJ}}(\%)$ & ${\rm Sig.}(\sigma)$ & $\mathcal{B}(\times 10^{-5})$\\[1mm]
            \hline
            $\chi_{c0}$ & $~284 \pm 44 $ & 3.05 & 5.6  & $3.51 \pm 0.54 \pm 0.29$  \\
            $\chi_{c1}$ & $~277 \pm 42 $ & 7.02 & 6.4  & $1.49 \pm 0.23 \pm 0.10$  \\
            $\chi_{c2}$ & $1038 \pm 56$ & 8.91 & 18 & $4.52 \pm 0.24 \pm 0.18$  \\[1mm]
            \hline \hline
        \end{tabular}
    \vspace{-0.2cm}
    \end{center}
\end{table}

\section{Summary}

In this talk, recently published analyses of BESIII are introduced briefly. BESIII has collected the largest data sample of about 10 billion $J/\psi$ and 2.7 billion $\psi(3686)$ in 2009, 2012 and 2021, which will definitely benefit not only the studies of charmonium decays but multiple fields including the transition between low-lying charmonium states ($\psi(3686) \to \eta_c/\eta_c(2S)$, $\psi(3686) \to \chi_{cJ}$...), the precise validation of the non-perturbative QCD calculation, the study of the light hadron spectroscopy, the search for the rare decays of charmonia (leptonic, semi-leptonic, invisible...) and the search for the new physics beyond standard model related with axion, dark matter etc.

\bibliographystyle{unsrt}

\begin{thebibliography}{99}

\bibitem{Kwong:1987mj}
W.~Kwong, J.~L.~Rosner and C.~Quigg,
\href{https://www.annualreviews.org/doi/10.1146/annurev.ns.37.120187.001545}{Ann. Rev. Nucl. Part. Sci. \textbf{37}, 325-382 (1987)}.

\bibitem{QuarkoniumWorkingGroup:2004kpm}
N.~Brambilla \textit{et al.} [Quarkonium Working Group],
[arXiv:\href{https://arxiv.org/abs/hep-ph/0412158}{hep-ph/0412158} [hep-ph]].

\bibitem{Eichten:2007qx}
E.~Eichten, S.~Godfrey, H.~Mahlke and J.~L.~Rosner,
\href{https://journals.aps.org/rmp/abstract/10.1103/RevModPhys.80.1161}{Rev. Mod. Phys. \textbf{80}, 1161-1193 (2008)}
[arXiv:\href{https://arxiv.org/abs/hep-ph/0701208}{hep-ph/0701208} [hep-ph]].

\bibitem{Brambilla:2010cs}
N.~Brambilla, S.~Eidelman, B.~K.~Heltsley, R.~Vogt, G.~T.~Bodwin, E.~Eichten, A.~D.~Frawley, A.~B.~Meyer, R.~E.~Mitchell and V.~Papadimitriou, \textit{et al.}
\href{https://link.springer.com/article/10.1140/epjc/s10052-010-1534-9}{Eur. Phys. J. C \textbf{71}, 1534 (2011)}
[arXiv:\href{https://arxiv.org/abs/1010.5827}{1010.5827} [hep-ph]].

\bibitem{Rosner:2011eg}
J.~L.~Rosner,
[arXiv:\href{https://arxiv.org/abs/1107.1273}{1107.1273} [hep-ph]].

\bibitem{Ablikim:2009aa}
M.~Ablikim \textit{et al.} [BESIII],
\href{https://www.sciencedirect.com/science/article/abs/pii/S0168900209023870?via\%3Dihub}{Nucl. Instrum. Meth. A \textbf{614}, 345-399 (2010)}
[arXiv:\href{https://arxiv.org/abs/0911.4960}{0911.4960} [physics.ins-det]].

\bibitem{Yu:IPAC2016-TUYA01}
C.~Yu, Z.~Duan, S.~Gu, Y.~Guo, X.~Huang, D.~Ji, H.~Ji, Y.~Jiao, Z.~Liu and Y.~Peng, \textit{et al.}
\href{https://accelconf.web.cern.ch/ipac2016/doi/JACoW-IPAC2016-TUYA01.html}{Proceedings of IPAC2016, Busan, Korea, 2016}.

\bibitem{Ablikim:2019hff}
M.~Ablikim \textit{et al.} [BESIII],
\href{https://iopscience.iop.org/article/10.1088/1674-1137/44/4/040001}{Chin. Phys. C \textbf{44}, no.4, 040001 (2020)}
[arXiv:\href{https://arxiv.org/abs/1912.05983}{1912.05983} [hep-ex]].

\bibitem{EcmsMea}
J.~Lu, Y.~Xiao, X.~Ji,
\href{https://link.springer.com/article/10.1007/s41605-020-00188-8}{Radiat. Detect. Technol. Methods {\bf 4}, 337–344 (2020)}.

\bibitem{EventFilter}
J.~W.~Zhang, L.~H.~Wu, S.~S.~Sun {\it et al.},
\href{https://link.springer.com/article/10.1007/s41605-022-00331-7}{Radiat. Detect. Technol. Methods {\bf 6}, 289–293 (2022)}.
  
\bibitem{etof1}
X.~Li {\it et al.}, \href{https://link.springer.com/article/10.1007/s41605-017-0014-2}{Radiat. Detect. Technol. Methods {\bf 1}, 13 (2017)}.

\bibitem{etof2}
Y.~X.~Guo {\it et al.}, \href{https://link.springer.com/article/10.1007/s41605-017-0012-4}{Radiat. Detect. Technol. Methods {\bf 1}, 15 (2017)}.

\bibitem{etof3}
P.~Cao, H.~F.~Chen, M.~M.~Chen, H.~L.~Dai, Y.~K.~Heng, X.~L.~Ji, X.~S.~Jiang, C.~Li, X.~Li and S.~B.~Liu, \textit{et al.}
\href{https://www.sciencedirect.com/science/article/abs/pii/S0168900219314068?via\%3Dihub}{Nucl. Instrum. Meth. A \textbf{953}, 163053 (2020)}.

\bibitem{BESIII:2023zcs}
M.~Ablikim \textit{et al.} [BESIII],
\href{https://link.springer.com/article/10.1007/JHEP05(2023)069}{JHEP \textbf{05}, 069 (2023)}
[arXiv:\href{https://arxiv.org/abs/2301.12922}{2301.12922} [hep-ex]].

\bibitem{Duncan:1980qd}
A.~Duncan and A.~H.~Mueller,
\href{https://www.sciencedirect.com/science/article/abs/pii/0370269380901082?via\%3Dihub}{Phys. Lett. B \textbf{93}, 119-124 (1980)}.

\bibitem{Jones:1981ff}
H.~F.~Jones and J.~Wyndham,
\href{https://www.sciencedirect.com/science/article/abs/pii/0550321382903972?via\%3Dihub}{Nucl. Phys. B \textbf{195}, 222-236 (1982)}.

\bibitem{Anselmino:1992rw}
M.~Anselmino and F.~Murgia,
\href{https://journals.aps.org/prd/abstract/10.1103/PhysRevD.47.3977}{Phys. Rev. D \textbf{47}, 3977-3983 (1993)}.

\bibitem{Zhou:2004mw}
H.~Q.~Zhou, R.~G.~Ping and B.~S.~Zou,
\href{https://www.sciencedirect.com/science/article/abs/pii/S0370269305002170?via\%3Dihub}{Phys. Lett. B \textbf{611}, 123-128 (2005)}
[arXiv:\href{https://arxiv.org/abs/hep-ph/0412221}{hep-ph/0412221} [hep-ph]].

\bibitem{Chernyak:1981zz}
V.~L.~Chernyak and A.~R.~Zhitnitsky,
\href{https://www.sciencedirect.com/science/article/abs/pii/055032138290445X?via\%3Dihub}{Nucl. Phys. B \textbf{201}, 492 (1982)}
[erratum: \href{https://www.sciencedirect.com/science/article/pii/0550321383902511?via\%3Dihub}{Nucl. Phys. B \textbf{214}, 547 (1983)}].

\bibitem{Yang:1950rg}
C.~N.~Yang,
\href{https://journals.aps.org/pr/abstract/10.1103/PhysRev.77.242}{Phys. Rev. \textbf{77}, 242-245 (1950)}.

\bibitem{BESII:2011hcd}
M.~Ablikim \textit{et al.} [BESII],
\href{https://journals.aps.org/prl/abstract/10.1103/PhysRevLett.107.092001}{Phys. Rev. Lett. \textbf{107}, 092001 (2011)}
[arXiv:\href{https://arxiv.org/abs/1104.5068}{1104.5068} [hep-ex]].

\bibitem{Huang:2021kfm}
Q.~Huang, J.~Z.~Wang, R.~G.~Ping and X.~Liu,
\href{https://journals.aps.org/prd/abstract/10.1103/PhysRevD.103.096006}{Phys. Rev. D \textbf{103}, no.9, 096006 (2021)}
[arXiv:\href{https://arxiv.org/abs/2102.07104}{2102.07104} [hep-ph]].

\bibitem{Chen:2013gka}
H.~Chen and R.~G.~Ping,
\href{https://journals.aps.org/prd/abstract/10.1103/PhysRevD.88.034025}{Phys. Rev. D \textbf{88}, no.3, 034025 (2013)}.

\bibitem{ParticleDataGroup:2022pth}
R.~L.~Workman \textit{et al.} [Particle Data Group],
\href{https://academic.oup.com/ptep/article/2022/8/083C01/6651666?login=false}{PTEP \textbf{2022}, 083C01 (2022)}.

\bibitem{BESIII:2020nme}
M.~Ablikim \textit{et al.} [BESIII],
\href{https://iopscience.iop.org/article/10.1088/1674-1137/44/4/040001}{Chin. Phys. C \textbf{44}, no.4, 040001 (2020)}
[arXiv:\href{https://arxiv.org/abs/1912.05983}{1912.05983} [hep-ex]].

\bibitem{BESIII:2022ksv}
M.~Ablikim \textit{et al.} [BESIII],
\href{https://journals.aps.org/prd/abstract/10.1103/PhysRevD.107.052007}{Phys. Rev. D \textbf{107}, no.5, 052007 (2023)}
[arXiv:\href{https://arxiv.org/abs/2211.11935}{2211.11935} [hep-ex]].

\bibitem{Appelquist:1974zd}
T.~Appelquist and H.~D.~Politzer,
\href{https://journals.aps.org/prl/abstract/10.1103/PhysRevLett.34.43}{Phys. Rev. Lett. \textbf{34}, 43 (1975)}.

\bibitem{Franklin:1983ve}
M.~E.~B.~Franklin, G.~J.~Feldman, G.~S.~Abrams, M.~S.~Alam, C.~A.~Blocker, A.~Blondel, A.~Boyarski, M.~Breidenbach, D.~L.~Burke and W.~C.~Carithers, \textit{et al.}
\href{https://journals.aps.org/prl/abstract/10.1103/PhysRevLett.51.963}{Phys. Rev. Lett. \textbf{51}, 963-966 (1983)}.

\bibitem{Chao:1996}
K. T. Chao, Y. F. Gu, and S. F. Tuan,
\href{https://iopscience.iop.org/article/10.1088/0253-6102/25/4/471}{Commun. Theor. Phys. 25, 471 (1996)}.

\bibitem{BESIII:2022hcv}
M.~Ablikim \textit{et al.} [BESIII],
\href{https://journals.aps.org/prd/abstract/10.1103/PhysRevD.106.032014}{Phys. Rev. D \textbf{106}, no.3, 032014 (2022)}
[arXiv:\href{https://arxiv.org/abs/2206.08807}{2206.08807} [hep-ex]].

\bibitem{BESIII:2022yoo}
M.~Ablikim \textit{et al.} [BESIII],
\href{https://journals.aps.org/prd/abstract/10.1103/PhysRevD.107.L091101}{Phys. Rev. D \textbf{107}, no.9, L091101 (2023)}
[arXiv:\href{https://arxiv.org/abs/2212.12165}{2212.12165} [hep-ex]].

\bibitem{Eichten:1978tg}
E.~Eichten, K.~Gottfried, T.~Kinoshita, K.~D.~Lane and T.~M.~Yan,
\href{https://journals.aps.org/prd/abstract/10.1103/PhysRevD.17.3090}{Phys. Rev. D \textbf{17}, 3090 (1978)}
[erratum: \href{https://journals.aps.org/prd/abstract/10.1103/PhysRevD.21.313.2}{Phys. Rev. D \textbf{21}, 313 (1980)}].

\bibitem{He:2008xb}
Z.~G.~He, Y.~Fan and K.~T.~Chao,
\href{https://journals.aps.org/prl/abstract/10.1103/PhysRevLett.101.112001}{Phys. Rev. Lett. \textbf{101}, 112001 (2008)}
[arXiv:\href{https://arxiv.org/abs/0802.1849}{0802.1849} [hep-ph]].

\bibitem{Voloshin:2005sd}
M.~B.~Voloshin,
\href{https://journals.aps.org/prd/abstract/10.1103/PhysRevD.71.114003}{Phys. Rev. D \textbf{71}, 114003 (2005)}
[arXiv:\href{https://arxiv.org/abs/hep-ph/0504197}{hep-ph/0504197} [hep-ph]].

\bibitem{Mo:2006cy}
X.~H.~Mo, C.~Z.~Yuan and P.~Wang,
\href{http://cpc.ihep.ac.cn/article/id/d7f266c1-9574-470c-9a0d-c9da20a5a2e2}{Chin. Phys. C \textbf{31}, 686-701 (2007)}
[arXiv:\href{https://arxiv.org/abs/hep-ph/0611214}{hep-ph/0611214} [hep-ph]].

\bibitem{Rosner:2004wy}
J.~L.~Rosner,
\href{https://www.sciencedirect.com/science/article/abs/pii/S0003491605000345?via\%3Dihub}{Annals Phys. \textbf{319}, 1-12 (2005)}
[arXiv:\href{https://arxiv.org/abs/hep-ph/0411003}{hep-ph/0411003} [hep-ph]].

\bibitem{Zhang:2009gy}
Y.~J.~Zhang and Q.~Zhao,
\href{https://journals.aps.org/prd/abstract/10.1103/PhysRevD.81.034011}{Phys. Rev. D \textbf{81}, 034011 (2010)}
[arXiv:\href{https://arxiv.org/abs/0911.5651}{0911.5651} [hep-ph]].

\bibitem{Ding:1991vu}
Y.~B.~Ding, D.~H.~Qin and K.~T.~Chao,
\href{https://journals.aps.org/prd/abstract/10.1103/PhysRevD.44.3562}{Phys. Rev. D \textbf{44}, 3562-3566 (1991)}.

\bibitem{Guo:2012tj}
Z.~k.~Guo, S.~Narison, J.~M.~Richard and Q.~Zhao,
\href{https://journals.aps.org/prd/abstract/10.1103/PhysRevD.85.114007}{Phys. Rev. D \textbf{85}, 114007 (2012)}
[arXiv:\href{https://arxiv.org/abs/1204.1448}{1204.1448} [hep-ph]].

\bibitem{Wang:2011yh}
Q.~Wang, X.~H.~Liu and Q.~Zhao,
\href{https://journals.aps.org/prd/abstract/10.1103/PhysRevD.84.014007}{Phys. Rev. D \textbf{84}, 014007 (2011)}
[arXiv:\href{https://arxiv.org/abs/1103.1095}{1103.1095} [hep-ph]].

\bibitem{Liu:2009dr}
X.~Liu, B.~Zhang and X.~Q.~Li,
\href{https://www.sciencedirect.com/science/article/pii/S0370269309004729?via\%3Dihub}{Phys. Lett. B \textbf{675}, 441-445 (2009)}
[arXiv:\href{https://arxiv.org/abs/0902.0480}{0902.0480} [hep-ph]].

\bibitem{Guo:2010ak}
F.~K.~Guo, C.~Hanhart, G.~Li, U.~G.~Meissner and Q.~Zhao,
\href{https://journals.aps.org/prd/abstract/10.1103/PhysRevD.83.034013}{Phys. Rev. D \textbf{83}, 034013 (2011)}
[arXiv:\href{https://arxiv.org/abs/1008.3632}{1008.3632} [hep-ph]].

\bibitem{CLEO:2005zky}
N.~E.~Adam \textit{et al.} [CLEO],
\href{https://journals.aps.org/prl/abstract/10.1103/PhysRevLett.96.082004}{Phys. Rev. Lett. \textbf{96}, 082004 (2006)}
[arXiv:\href{https://arxiv.org/abs/hep-ex/0508023}{hep-ex/0508023} [hep-ex]].

\bibitem{Zhang:2009kr}
Y.~J.~Zhang, G.~Li and Q.~Zhao,
\href{https://journals.aps.org/prl/abstract/10.1103/PhysRevLett.102.172001}{Phys. Rev. Lett. \textbf{102}, 172001 (2009)}
[arXiv:\href{https://arxiv.org/abs/0902.1300}{0902.1300} [hep-ph]].

\bibitem{Li:2013zcr}
G.~Li, X.~h.~Liu, Q.~Wang and Q.~Zhao,
\href{https://journals.aps.org/prd/abstract/10.1103/PhysRevD.88.014010}{Phys. Rev. D \textbf{88}, no.1, 014010 (2013)}
[arXiv:\href{https://arxiv.org/abs/1302.1745}{1302.1745} [hep-ph]].

\bibitem{Anwar:2016mxo}
M.~N.~Anwar, Y.~Lu and B.~S.~Zou,
\href{https://journals.aps.org/prd/abstract/10.1103/PhysRevD.95.114031}{Phys. Rev. D \textbf{95}, no.11, 114031 (2017)}
[arXiv:\href{https://arxiv.org/abs/1612.05396}{1612.05396} [hep-ph]].

\bibitem{Kuraev:1985hb}
E.~A.~Kuraev and V.~S.~Fadin,
Sov. J. Nucl. Phys. \textbf{41}, 466-472 (1985).

\bibitem{WorkingGrouponRadiativeCorrections:2010bjp}
S.~Actis \textit{et al.} [Working Group on Radiative Corrections and Monte Carlo Generators for Low Energies],
\href{https://link.springer.com/article/10.1140/epjc/s10052-010-1251-4}{Eur. Phys. J. C \textbf{66}, 585-686 (2010)}
[arXiv:\href{https://arxiv.org/abs/0912.0749}{0912.0749} [hep-ph]].

\bibitem{Sun:2020ehv}
W.~Sun, T.~Liu, M.~Jing, L.~Wang, B.~Zhong and W.~Song,
\href{https://link.springer.com/article/10.1007/s11467-021-1085-6}{Front. Phys. (Beijing) \textbf{16}, no.6, 64501 (2021)}
[arXiv:\href{https://arxiv.org/abs/2011.07889}{2011.07889} [hep-ex]].

\bibitem{BESIII:2020bgb}
M.~Ablikim \textit{et al.} [BESIII],
\href{https://journals.aps.org/prd/abstract/10.1103/PhysRevD.102.031101}{Phys. Rev. D \textbf{102}, no.3, 031101 (2020)}
[arXiv:\href{https://arxiv.org/abs/2003.03705}{2003.03705} [hep-ex]].

\bibitem{BESIII:2023bnk}
M.~Ablikim \textit{et al.} [BESIII],
\href{https://journals.aps.org/prd/abstract/10.1103/PhysRevD.107.092004}{Phys. Rev. D \textbf{107}, no.9, 092004 (2023)}
[arXiv:\href{https://arxiv.org/abs/2302.12579}{2302.12579} [hep-ex]].

\bibitem{DASP:1975xwv}
W.~Braunschweig \textit{et al.} [DASP],
\href{https://www.sciencedirect.com/science/article/abs/pii/0370269375904827?via\%3Dihub}{Phys. Lett. B \textbf{57}, 407-412 (1975)}.

\bibitem{Feldman:1975bq}
G.~J.~Feldman, B.~Jean-Marie, B.~Sadoulet, F.~Vannucci, G.~S.~Abrams, A.~Boyarski, M.~Breidenbach, F.~Bulos, W.~Chinowsky and C.~E.~Friedberg, \textit{et al.}
\href{https://journals.aps.org/prl/abstract/10.1103/PhysRevLett.35.821}{Phys. Rev. Lett. \textbf{35}, 821 (1975)}
[erratum: \href{https://journals.aps.org/prl/abstract/10.1103/PhysRevLett.35.1184.3}{Phys. Rev. Lett. \textbf{35}, 1184 (1975)}].

\bibitem{Wong:1999hc}
S.~M.~H.~Wong,
\href{https://link.springer.com/article/10.1007/s100520000376}{Eur. Phys. J. C \textbf{14}, 643-671 (2000)}
[arXiv:\href{https://arxiv.org/abs/hep-ph/9903236}{hep-ph/9903236} [hep-ph]].

\bibitem{Liu:2009vv}
X.~H.~Liu and Q.~Zhao,
\href{https://journals.aps.org/prd/abstract/10.1103/PhysRevD.81.014017}{Phys. Rev. D \textbf{81}, 014017 (2010)}
[arXiv:\href{https://arxiv.org/abs/0912.1508}{0912.1508} [hep-ph]].

\bibitem{BESIII:2012wgp}
M.~Ablikim \textit{et al.} [BESIII],
\href{https://journals.aps.org/prd/abstract/10.1103/PhysRevD.86.052004}{Phys. Rev. D \textbf{86}, 052004 (2012)}
[arXiv:\href{https://arxiv.org/abs/1207.5646}{1207.5646} [hep-ex]].

\bibitem{Achasov:2023gey}
M.~Achasov, X.~C.~Ai, R.~Aliberti, Q.~An, X.~Z.~Bai, Y.~Bai, O.~Bakina, A.~Barnyakov, V.~Blinov and V.~Bobrovnikov, \textit{et al.}
[arXiv:\href{https://arxiv.org/abs/2303.15790}{2303.15790} [hep-ex]].

\end{thebibliography}

\end{document}